\documentclass[10pt,aps,pra,reprint,showpacs,amsmath,amssymb]{revtex4-1}
\usepackage[latin9]{inputenc}
\usepackage{amsmath}
\usepackage[pdftex]{graphicx}
\usepackage{esint}

\makeatletter
\@ifundefined{textcolor}{}
{%
 \definecolor{BLACK}{gray}{0}
 \definecolor{WHITE}{gray}{1}
 \definecolor{RED}{rgb}{1,0,0}
 \definecolor{GREEN}{rgb}{0,1,0}
 \definecolor{BLUE}{rgb}{0,0,1}
 \definecolor{CYAN}{cmyk}{1,0,0,0}
 \definecolor{MAGENTA}{cmyk}{0,1,0,0}
 \definecolor{YELLOW}{cmyk}{0,0,1,0}
}

\usepackage[pdftex]{graphicx}% Include figure files
\usepackage{float}

 \usepackage{setspace}

\makeatother

\begin{document}

\title{Comparison of strong-coupling theories for a two-dimensional Fermi
gas}

\author{Brendan C. Mulkerin}

\author{Kristian Fenech}

\author{Paul Dyke}

\author{Chris J. Vale}

\author{Xia-Ji Liu}

\author{Hui Hu}

\affiliation{Centre for Quantum and Optical Science, Swinburne University of Technology,
Melbourne 3122, Australia.}

\date{\today}
\begin{abstract}
Understanding the formation of Cooper pairs and the resulting thermodynamic
properties of low-dimensional Fermi gases is an important area of
research, which may help build our understanding of 
other low-dimensional systems such as high temperature superconductors.
In lower dimensions quantum fluctuations are expected to play an increasingly
important role and the reliability of strong-coupling theories becomes
questionable. Here, we present a comparison of recent thermodynamic
measurements and theoretical predictions from different many-body
$T$-matrix theories for a two-dimensional strongly interacting Fermi
gas in the normal state. We find that the fully self-consistent $T$-matrix
theory provides the best description of the experimental data over
a wide range of temperatures and interatomic interactions. %Our comparison
%reveals the crucial role played by the interactions between Cooper
%pairs and suggests that the future development of a quantitative strong-coupling
%theory for two-dimensional Fermi superfluids must explicitly take
%into account the diagrams that are responsible for pair-pair interactions. 
\end{abstract}

\pacs{05.30.Fk, 67.85.-d, 03.75.Hh}

\maketitle
The understanding of pairing of fermions in strongly interacting two-dimensional
(2D) Fermi gases is of great interest to condensed matter physics,
where the pairing mechanism in high-temperature superconductors remains
elusive \cite{Wen2006}. In order to theoretically understand these
systems new approaches are required to treat strong interactions as
one encounters a ``strongly correlated'' regime. 
%Such systems occur in many fundamental systems ranging from strongly-interacting electrons, neutron stars and 
%quark-gluon plasmas \cite{things}.

The main theoretical difficulty in describing strongly interacting
systems is the absence of any small-coupling parameter, which is crucial
for truncating perturbative approaches. Due to large quantum fluctuations,
mean-field theories do not describe the strongly correlated Fermi
gas away from $T=0$ \cite{loktev2001,Levinsen2015}, where correlations beyond the
single-particle picture play an important role. There are numerous
efforts to develop strong-coupling perturbation theories in both two
and three dimensions, notably many-body $T$-matrix fluctuation theories
\cite{Haussmann1993,Bauer2014,Marsilgio15,*Pietil2012,*watanabe,Hu2007,*Hu2010,*ohashi2009},
however, the accuracy of such methods is not well understood. Sophisticated
quantum Monte Carlo (QMC) simulations have been developed in solving
strongly coupled systems, such as diffusion Monte Carlo \cite{Bertaina2011},
auxiliary field Monte Carlo \cite{Shi2015}, lattice Monte Carlo \cite{anderson2015},
and diagrammatic quantum Monte Carlo \cite{Vlietinck2014}, however,
these approaches also have difficulty evaluating the equation
of state. The virial expansion has also been studied in harmonically
trapped \cite{Liu20102D} and homogeneous systems \cite{Barth2014,Ngamp13},
giving exact results in the high-temperature
limit.

Recent developments in the experimental realization of two-dimensional
ultracold Fermi gases with a tunable interaction through Feshbach resonances,
densities, and temperatures provide a unique opportunity to understand and benchmark strong-coupling theories 
for the two-dimensional BEC-BCS crossover \cite{Dyke2014,murthy2014,fenech2015} and the Berezinskii-Kosterlitz-Thouless (BKT) transition
\cite{Ries2014,*Murthy2015}. In these experiments it is possible
to extract the density versus chemical potential at a fixed interaction
 directly from the measured density profile
in the trap \cite{Ku2012}, allowing a direct comparison between theoretical
and experimental results. Pairing and superfluidity have been
studied for two-dimensional ultracold gases \cite{frolich2011,*feld2011}, where the formation of pairs
above the superfluid transition $T_{c}$, the pseudogap, was
examined. The formation of pairs above $T_c$ is a precursor to superfluidity
and is important in understanding the BKT transition. In two dimensions
the pseudogap regime is expected to be more pronounced than in three-dimensional
systems due to the increasingly important quantum fluctuations in
low-dimensions \cite{Perali2011,*Gaebler2010,*Stewart2008}.

In this paper we draw upon recent experimental data
as a benchmark and present a direct comparison of several $T$-matrix
theories as has been performed in 3D \cite{Huicomparitive08}. Examining
the thermodynamic properties of the density equation of state, pressure
equation of state, and compressibility, we show that the fully self-consistent
theory successfully describes a 2D Fermi gas over a broad range of
temperatures and interaction strengths. We compute the spectral function
of the 2D Fermi gas for a fixed interaction strength and temperature
currently available to experiment and compare the onset of a pseudogap
from the $T$-matrix theories. This contrasts with the 3D case where 
the strong-coupling theories disagree over
for the existence of a pseudogap \cite{Zweger2009,*Chien2010}. 
%Our results should be useful in applying different $T$-matrix approximations
%to high-$T_{c}$ superconductors and other strongly-interacting low-dimensional
%systems.

The theoretical models compared in this paper are three $T$-matrix approximations, 
described briefly here, and for a more detailed description we refer to 
Refs.~\cite{loktev2001,Huicomparitive08,Haussmann93}. The $T$-matrix theories involve a partial
summation of the infinite set of ladder diagrams, which are generally
accepted as the most important contribution in strongly interacting
systems. We wish to study the properties of the normal state of a
2D system, i.e., above $T_{c}$, where we will set $\hbar=1$, $k_{{\rm B}}=1$, the mass 
$2M=1$, and keep dimensionful variables where
instructive. The dressed Green's function is given by Dyson's equation,
\begin{alignat}{1}
G^{-1}(\mathbf{k},i\omega)=G_{0}^{-1}(\mathbf{k},i\omega)-\Sigma(\mathbf{k},i\omega),\label{Eq:Dress_Green}
\end{alignat}
where $\omega=(2m+1)\pi/\beta$ for integer $m$, $\beta=1/T$, $\Sigma(\mathbf{k},i\omega)$
is the self-energy, and the free Green's function is given by $G_{0}(\mathbf{k},i\omega)^{-1}=(i\omega-\varepsilon_{\mathbf{k}}+\mu)$
with $\varepsilon_{\mathbf{k}}=\mathbf{k}^{2}/(2M)$. The self-energy is given in real space
as 
\begin{alignat}{1}
\Sigma(\mathbf{x},\tau)=G(-\mathbf{x},-\tau)\Gamma(\mathbf{x},\tau),\label{Eq:self_ener}
\end{alignat}
where the regularized vertex function is given through the Bethe-Salpeter
equations
\begin{alignat}{1}
\Gamma(\mathbf{K},i\Omega)=\frac{1}{g_{0}^{-1}(\Lambda)+\chi(\mathbf{K},i\Omega)}.\label{Eq:Gamma}
\end{alignat}
Here, $\Omega=2n\pi/\beta$ are the bosonic Matsubara frequencies for
integer $n$, and the pair propagator is given as 
\begin{alignat}{1}
\chi(\mathbf{K},i\Omega)=\int\frac{d\mathbf{k}}{(2\pi)^{2}}\frac{1}{\beta}\sum_{\omega}G(\mathbf{K}-\mathbf{k},i\Omega-i\omega)G(\mathbf{k},i\omega).\label{Eq:pair_prop}
\end{alignat}
The coupling term $g_{0}^{-1}(\Lambda)$ is expressed in terms of
the physical binding energy, $\varepsilon_{B}=\hbar^{2}/(Ma_{\rm 2D}^{2})$, which is always
present in a 2D Fermi gas \cite{Adhikari1986},
and $a_{\rm 2D}$ is the $s$-wave scattering length in 2D, 
\begin{alignat}{1}
g_{0}^{-1}(\Lambda)= - \int^{\Lambda}\frac{d\mathbf{k}}{(2\pi)^{2}}\frac{1}{2\varepsilon_{\mathbf{k}}+\varepsilon_{B}}.
\end{alignat}
Equations \eqref{Eq:Dress_Green}$-$\eqref{Eq:pair_prop}, with the
regularized two-body interaction, constitute a self-consistent set
of coupled integral equations which we solve on a logarithmic grid
until convergence is reached. We solve the set of integral equations
for a fixed temperature $T> T_{c}$ and fixed coupling constant
$\smash{\eta=-1/2\ln(\varepsilon_{B}/2E_{\rm F})=\ln(k_{\rm F}a_{2D})}$, where
the Fermi energy is given by $E_{\rm F}=k_{\rm F}^2/2M$
and $k_{\rm F}$ is the Fermi momentum. Here, $T_c$ is defined by the 
divergence of the $T$ matrix, the Thouless criterion, $\Gamma^{-1}(\mathbf{q}=0,\Omega=0)\left. \right |_{T=T_c}=0$.	 
In two dimensions the $T$-matrix approximation does not recover the BKT transition 
and the transition temperature is found to be $T_c = 0$, 
thus we restrict ourselves to an analysis away from the superfluid transition \cite{loktev2001}. The chemical
potential $\mu$ is a free parameter and is fixed by the number equation
$n=-2G(\mathbf{x}=0,\tau=0^{-})$. As in 3D we need to calculate
the Fourier transforms efficiently and precisely, carefully considering
the singular behavior of the functions $G(\mathbf{x},\tau)$, $\Gamma(\mathbf{x},\tau)$,
and $\Sigma(\mathbf{x},\tau)$ and their logarithmic divergences.

From the general self-consistent set of equations it is possible to
choose the different $T$-matrix schemes based upon the choice of interacting
and free Green's functions. Firstly, we have the simplest method,
the \textrm{NSR} theory, which was originally used to calculate
the thermodynamic potential by Nozières and Schmitt-Rink for the BEC-BCS
crossover \cite{nozieres1985bose}. This theory was extended to two
dimensions \cite{randeria1989bound,*randeria1990superconductivity,*engelbrecht1990new,*schmitt1989pairing}
and is equivalent to calculating a truncated self-energy within the
Dyson expansion \cite{serene1989}, $G=G_{0}+G_{0}\Sigma G_{0}$.
The \textrm{NSR} theory can be extended to include all repeated scatterings by summing
the full series in the Dyson equation and has been extensively studied
in the literature \cite{Marsilgio15,Pietil2012,watanabe}, in this
work we will only consider the initial truncated case.  %We refer to this scheme as
%the \textit{NSR} theory, equivalently it can be referred to as $G_{0}G_{0}$
%owing to the presence of two bare fermion propagators, $G_{0}$ in
%the vertex function of Eq.~\eqref{Eq:Gamma}.

The second $T$-matrix theory considered is the $GG_{0}$ theory, where
$G$ is an interacting or dressed Green's function. This elevated
Green's function must be calculated self-consistently, adding considerable
time to the computation. One bare $G_{0}$ and one self-consistent
$G$ enter the vertex equation, while there is a bare $G_{0}$
kept in the definition of the fermionic self-energy, Eq.~\eqref{Eq:self_ener}.

The final $T$-matrix theory studied in this paper is the $GG$ theory,
where all the single particle Green's function in the vertex and self-energy
have been self-consistently calculated. The $GG$ scheme has been
studied extensively in the literature for three dimensions
and recently in two dimensions and is known as the Luttinger-Ward theory \cite{Bauer2014}. In three
dimensions the $GG$ $T$ matrix yields the best results for calculating the thermodynamic properties
of the unitary gas compared to experiment and quantum Monte Carlo \cite{Ku2012}. However,
the $GG$ theory is far from being exact and has its own shortcomings.
For two dimensions in the dilute BEC limit the $GG$ 
theory is unphysical as it predicts a constant interaction between
composite bosons. The non-self- consistent $T$-matrix theories do not contain this unphysical behavior and better
describe the deep BEC limit \cite{he2015}. Computationally, the non-self consistent calculations are the
simplest to find a converged solution, and it is instructive to compare the $T$-matrix theories
to experiment.

\begin{figure}
\includegraphics[width=0.9\columnwidth]{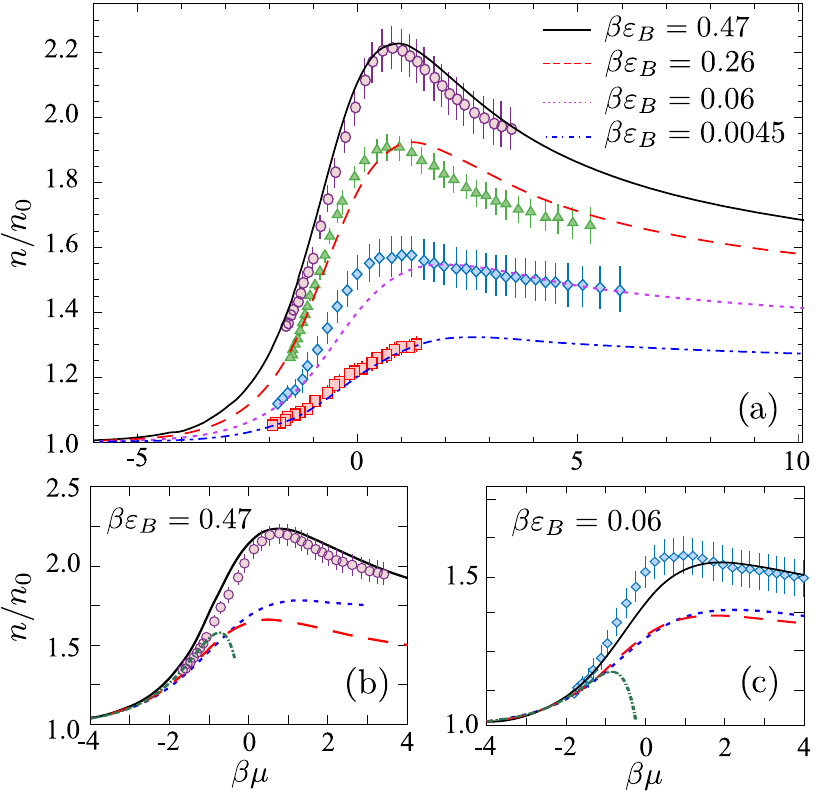} \protect\caption{(Color online) (a) The density equation of state, $n/n_{0}$ normalized
by an ideal system at the same temperature for the
$GG$ theory and experimental results at interaction strengths $\beta\varepsilon_{B}=0.47$
(black solid and purple circles), $\beta\varepsilon_{B}=0.26$ (red
dashed and green triangles), $\beta\varepsilon_{B}=0.06$ (purple
dotted and blue diamonds) and $\beta\varepsilon_{B}=0.0045$ (blue
dotted-dashed and red squares). Figures (b) and
(c) show a comparison of the three $T$-matrix theories $GG$ (black
solid), $GG_{0}$ (red dashed), \textrm{NSR} (blue dotted), third-order viral expansion 
(green dotted-dashed) and experiment for interaction strengths $\beta\varepsilon_{B}=0.47$ and $\beta\varepsilon_{B}=0.06$.}

\label{fig:denseos} 
\end{figure}

From the converged Green's functions we can find the density equation of state 
 as a function of $\beta\mu$
for a fixed interaction strength $\beta\varepsilon_{B}$. The density
equation of state is given in Fig.~\ref{fig:denseos}(a) plotted
as a function $\beta\mu$  for interaction strengths $\beta\varepsilon_{B}=0.47$
(black solid and purple circles), $\beta\varepsilon_{B}=0.26$ (red
dashed and green triangles), $\beta\varepsilon_{B}=0.06$ (purple
dotted and blue diamonds), and $\beta\varepsilon_{B}=0.0045$ (blue
dotted-dashed and red squares) for the self-consistent $GG$ theory and experiment.  
 To expose the effects of interactions
we have normalized the densities by that of an ideal Fermi gas at
the same temperature, $n_{0}=2\ln\left[1+e^{\beta\mu}\right]/\lambda_{{\rm T}}^{2}$,
where $\lambda_{{\rm T}}^{2}=2\pi/T$ is the thermal wavelength. 

The experimental data shown are taken from Ref.~\cite{fenech2015},
and we briefly describe the experiment here. An isolated 2D Fermi gases of $^6$Li atoms
 is produced in the lowest two spin states $|F = 1/2, m_F = \pm 1/2\rangle$. 
The cloud is confined to a blue-detuned TEM$_{01}$ mode laser beam that provides tight 
confinement along $z$ with $\omega_z / 2 \pi = 5.15$ kHz. Radial confinement is provided 
by a residual magnetic field curvature when the Feshbach coils are applied and produces a 
radially symmetric potential with $\omega_r / 2 \pi = 26$ Hz. The clouds are prepared in the 
kinematically 2D regime \cite{Dyke2014} where $N \approx 16000$ [$=0.4N_{2D}^{(Id.)}$, where 
$N_{2D}^{(Id.)} = (\omega_z/\omega_r)^2$ is the critical atom number] 
with a temperature range of 20-60 nK.

\begin{table}[t!]
\begin{center}
%S[table-format=3.2]%
\begin{tabular}{l@{\extracolsep{10pt} }llll}
\hline\hline
$\beta \varepsilon_B$ & $B$~(G) & $a_\mathrm{3D}$~$(a_0)$ & $\varepsilon_B$~(Hz) \\
\hline
0.0045 & 972 & $\,\,\,-4618.6$   & $\,\,\,\,\,\, 4.24 $  \\
0.06   & 920 & $\,\,\,-6354.2$   & $\,\,\, 21.00$  \\
0.26   & 880 & $-10289.6$  			 & 106.99 \\
0.47   & 865 & $-14249.9$        & 222.18 \\
\hline\hline
\end{tabular}
\caption{Values of the magnetic fields, scattering length $a_{\rm 3D}$, and binding energy $\varepsilon_B$ for the values of $\beta \varepsilon_B$ reported in Fig.~\ref{fig:denseos}.}
\label{table1}
\end{center}
\end{table}

Imaging of the cloud takes place along $z$ to directly obtain the density $n(x,y)$. 
Due to the cylindrically symmetric harmonic trap $V_r(x,y)$ we can azimuthally average the images to obtain $n(V_r)$. 
The $n(V_r)$ data is then used to construct a model independent equation of state 
for the dimensionless compressibility $\tilde{\kappa} = \kappa/\kappa_0$ and dimensionless 
pressure $\tilde{p} = P/P_0$ analogous to Refs.~\cite{Ku2012, dali14}, where $P_{0}=n_0E_{F}/2$ and $\kappa_{0}=1/(n_0E_{{\rm F}})$,
 at each magnetic field shown in 
Table \ref{table1}. From these dimensionless values one can obtain the density equation of state, 
where the reader is referred to Ref.~\cite{fenech2015} for more of the experimental details.
 Comparing the $GG$ theory and experimental results,
there is good agreement for all the interactions across a broad range
of temperatures. 

\begin{figure}
\includegraphics{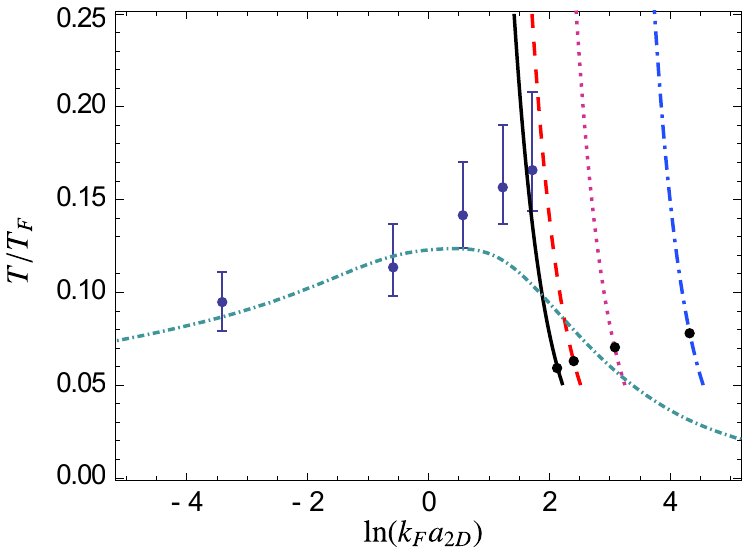} \protect\caption{(Color online) (a) 
Constant curves of $\beta\varepsilon_B$ given by $\beta\varepsilon_{B}=0.47$
(black solid), $\beta\varepsilon_{B}=0.26$ (red
dashed), $\beta\varepsilon_{B}=0.06$ (purple
dotted) and $\beta\varepsilon_{B}=0.0045$ (blue
dot-dashed). The black dots on each curve are for $\beta\mu=10$. The experimentally determined BKT transition
is given by blue dots with their respective error from Ref.~\cite{Murthy2015}, and the theoretical values from
Ref.~\cite{Bighin2015} are also given (green short-dotted-dashed).}

\label{fig:Tclogkfa} 
\end{figure}

In Figs.~\ref{fig:denseos}(b) and \ref{fig:denseos}(c) we compare the density equation
of state for an interaction strength of $\beta\varepsilon_{B}=0.47$
and $\beta\varepsilon_{B}=0.06$, respectively, from each $T$-matrix theory
$GG$ (black solid), $GG_{0}$ (blue dotted), \textrm{NSR} (red dashed), 
third-order virial expansion (green dotted-dashed) and their
equivalent experimental results shown with purple circles and blue diamonds.
The behavior of the density equation of state for the $GG_{0}$ and
\textrm{NSR} theories is qualitatively the same, however the results
are considerably different from the $GG$ theory and experiment. We see that the
$GG_{0}$ and \textrm{NSR} theories significantly under estimate the
density in the strongly interacting regime and in the low temperature,
weakly interacting regime. The \textrm{NSR}
results in Fig.~\ref{fig:denseos}(b) finish at a temperature of $T/T_{{\rm F}}\approx0.19$, where $T_{{\rm F}}$
is the Fermi temperature,
due to the reliability of the procedure and where the inverse of the vertex function
is close to zero \cite{Marsilgio15}.

 As we go from the high-temperature regime, $\beta\mu=-\infty$ to lower
temperatures the gas exhibits a maximum around $\beta\mu\simeq1$,
implying that interactions are strongest at intermediate temperatures.
This is understood from the interaction strength $\beta\varepsilon_{B}$,
for a decreasing temperature, $T/T_{{\rm F}}$, 
 corresponds to an increasing interaction $\eta  = \ln(k_{{\rm F}}a_{{\rm 2D}})$.
We see in the low-temperature regime the system is becoming
a weakly interacting gas. This behavior can be seen as we plot constant curves of $\beta\varepsilon_B$
for $T/T_{\rm F}$ as a function of $\ln(k_{\rm F}a_{\rm 2D})$ 
 in Fig.~\ref{fig:Tclogkfa}, where the curves are given by $\beta\varepsilon_{B}=0.47$
(black solid), $\beta\varepsilon_{B}=0.26$ (red
dashed), $\beta\varepsilon_{B}=0.06$ (purple
dotted), and $\beta\varepsilon_{B}=0.0045$ (blue
dotted-dashed). The black dots correspond to a value of $\beta\mu=10$. For comparison, we have
 plotted the experimentally determined BKT 
transition temperature from Ref.~\cite{Murthy2015} and the most recent calculation from
Ref.~\cite{Bighin2015}, where they have calculated the superfluid transition for the BEC-BCS crossover.

\begin{figure}
\includegraphics{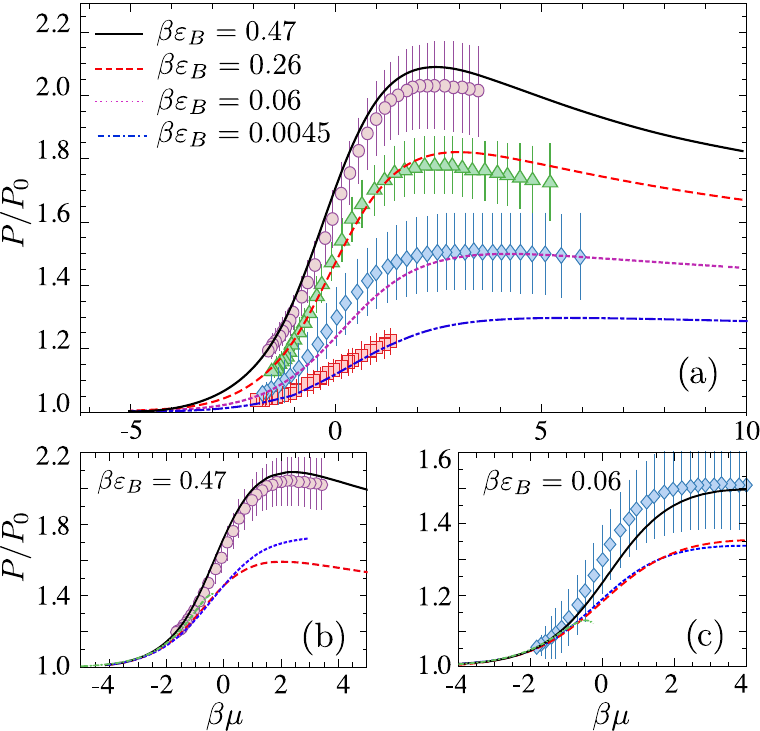} \protect\caption{(Color online) (a) 
The pressure equation of state $P/P_{0}$ normalized
by an ideal system at the same temperature for the
$GG$ theory and experimental results at interaction strengths $\beta\varepsilon_{B}=0.47$
(black solid and purple circles), $\beta\varepsilon_{B}=0.26$ (red
dashed and green triangles), $\beta\varepsilon_{B}=0.06$ (purple
dotted and blue diamonds), and $\beta\varepsilon_{B}=0.0045$ (blue
dotted-dashed and red squares). Figures (b) and (c) show a comparison
of the three $T$-matrix theories and experiment $GG$ (black solid),
$GG_{0}$ (red dashed), and \textrm{NSR} (blue dotted) for interaction
strengths $\beta\varepsilon_{B}=0.47$ and $\beta\varepsilon_{B}=0.06$.}

\label{fig:presseos} 
\end{figure}

From the density equation of state we can find the pressure through the Gibbs-Duhem relation 
\begin{alignat}{1}
P(\mu)\lambda_{T}^{4}=\int_{-\infty}^{\beta\mu}n(\beta\mu')\lambda_{T}^{2}d(\beta\mu').
\end{alignat}
In order to accurately calculate the lower limit of the integration
we have used the virial expansion to second order \cite{Barth2014,*Ngamp13}
for values of the density as $\beta\mu\rightarrow-\infty$. In Fig.~\ref{fig:presseos}(a)
we plot the normalised pressure as a function of $\beta\mu$ for interaction
strengths $\beta\varepsilon_{B}=0.47$ (black solid), $\beta\varepsilon_{B}=0.26$
(red dashed), $\beta\varepsilon_{B}=0.06$ (purple dotted), and $\beta\varepsilon_{B}=0.0045$
(blue dotted-dashed) for the self-consistent $GG$ theory. We have normalized
the pressure by that of an ideal Fermi gas at the same temperature,
$P_{0}\lambda_{T}^{4}=-2\pi{\rm Li}_{2}\left(-e^{\beta\mu}\right)$
and ${\rm Li}$ is the polylogarithm. The experimental data in Fig.~\ref{fig:presseos}(a)
are shown for the same interaction strengths as the theoretical
results, $\beta\varepsilon_{B}=0.47$ (purple circles), $\beta\varepsilon_{B}=0.26$
(green triangles), $\beta\varepsilon_{B}=0.06$ (blue diamonds) and
$\beta\varepsilon_{B}=0.0045$ (red squares), allowing for direct
comparison. We see that
there is good agreement for the four interactions across a broad set
of temperatures, showing the maximum in the pressure equation of state
near $\beta\mu\simeq1$ where the interactions are strongest. 

\begin{figure}
\includegraphics[width=0.9\columnwidth]{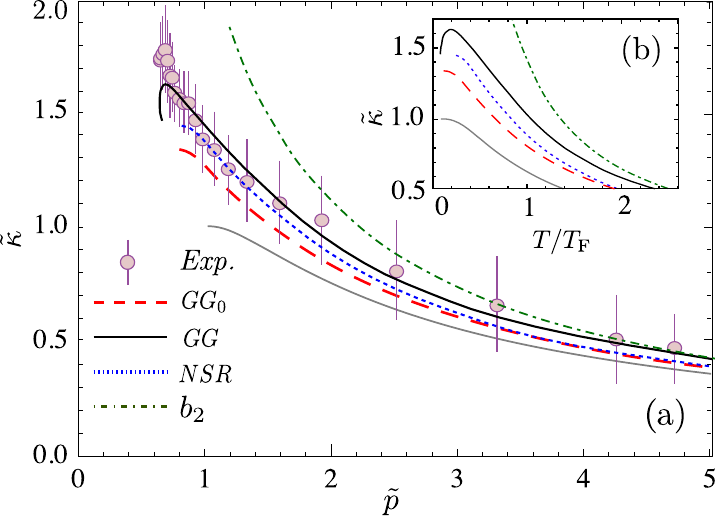} \protect\caption{(Color online) (a) 
The compressibility $\tilde{\kappa}=\kappa/\kappa_{0}$ plotted as a function of the
 pressure $\tilde{p}=P/P_{0}$ and (b) compressibility as a function of reduced temperature,
where $P_{0}=n_0E_{{\rm F}}/2$ and $\kappa_{0}=1/(n_0E_{{\rm F}})$ are
the ideal compressibility and pressure at zero temperature and interaction
of $\beta\varepsilon_{B}=0.47$. In each plot we have the $T$-matrix
theories $GG$ (black solid), $GG_{0}$ (red dashed), \textrm{NSR}
(blue dotted), and experiment (purple circles). For comparison, we plot also the prediction from the second-order
 virial expansion (green dotted-dashed) and ideal compressibility (gray) as a function of pressure or temperature. }

\label{fig:pressvcom} 
\end{figure}

In Figs.~\ref{fig:presseos}(b) and (c) we plot the pressure equation
of state for interaction strengths of $\beta\varepsilon_{B}=0.47$
and $\beta\varepsilon_{B}=0.06$, respectively, for the three $T$-matrix
theories $GG$ (black solid), $GG_{0}$ (blue dotted), \textrm{NSR}
(red dashed), and compare to the experimental results. The $GG_{0}$
and \textrm{NSR} underestimate the pressure in the strongly interacting
regime compared to the $GG$ theory, as we have seen in the density equation of state.% Comparing to the experimental results it is clear in the strongly interacting regime the $GG_0$ and {\it NSR} schemes considerably underestimate the correlations in the system.

The compressibility can be found from the density equation of state
through the relation 
\begin{alignat}{1}
\kappa=\frac{\beta}{n^{2}}\frac{\partial n}{\partial(\beta\mu)}\bigg|_{T}=\frac{\lambda_{T}^{4}}{2\pi}\frac{1}{(n\lambda_{T}^{2})^{2}}\frac{\partial n\lambda_{T}^{2}}{\partial(\beta\mu)}\bigg|_{T},
\end{alignat}
where we have written the dimensionless form for clarity. We plot
the compressibility $\tilde{\kappa}=\kappa/\kappa_{0}$ 
as a function of pressure $\tilde{p}=P/P_{0}$, normalized with their
ideal values at zero temperature,
in Fig.~\ref{fig:pressvcom}(a). 
%This data can be extracted directly
%from the experimental results without fitting to a model, such as
%the virial expansion. 
From the universal function $\tilde{\kappa}(\tilde{p})$
several other thermodynamic properties of the experimental system can
be found \cite{Ku2012,fenech2015}. Looking at Fig.~\ref{fig:pressvcom}(a), we see the compressibility
for all three $T$-matrix theories rises above that of the ideal gas with
the $GG$ $T$-matrix decreasing at lower pressure. We expect the 
the normalized compressibility for the three $T$-matrix
theories to lower, which in 3D marks the onset of
 pair formation and superfluidity \cite{Ku2012}. However, 
for an interaction of $\beta\varepsilon_B=0.47$, only the fully 
self-consistent $GG$ theory is reliable at low temperatures 
and the lowering is not seen in the $GG_0$ and {\rm NSR} theories. We see
all three theories are similar to the experimental results,
with the {\rm NSR} theory matching well for low pressures, and the
 $GG$ theory closely matches  for
a wide range of values, as we would expect from the similarity
found in the density and pressure equation of state for $\beta\varepsilon_{B}=0.47$.
At low pressure the $GG$ curve does not reach the same maximum as
the experimental results and begins to lower. It is difficult to assess whether the
experimental data show a similar feature due to the noise in the data;
lower temperatures would be required for a further comparison.
%With current achievable temperatures and resolution in experiment
%it is difficult to determine if the compressibility is in fact lowering. 
%, which in 3D marks the onset of
%

In Fig.~\ref{fig:pressvcom}(b) we plot the scaled compressibility
as a function of reduced temperature. We explicitly see here for low
temperatures the $GG$ theory decreasing from a maximum value for
temperatures below $T\simeq0.2T_{{\rm F}}$. %The small upturn in the $GG_{0}$
%plot indicates where the calculation breaks down and converges to unphysical solutions.

\begin{figure}
\includegraphics[width=0.9\columnwidth]{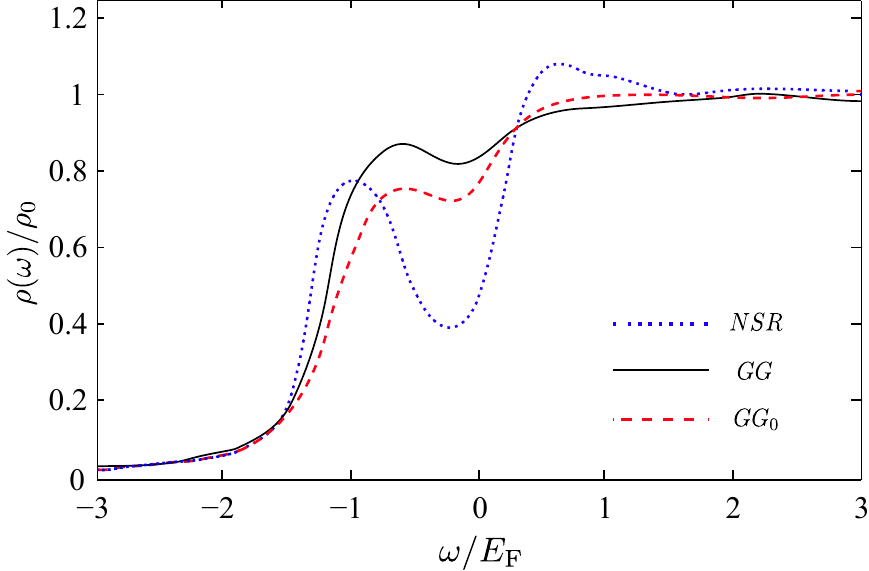} \protect\caption{(Color online) The density of states $\rho(\omega)$ is shown in
units of the non-interacting density of states at the Fermi surface $\rho_{0}=M/2\pi$,
for $\beta\varepsilon_{B}=0.47$ and $T/T_{{\rm F}}=0.2$ for $GG$
(black solid), $GG_{0}$ (red dashed), and \textrm{NSR} (blue dotted).}

\label{fig:dos} 
\end{figure}

For the three $T$-matrix calculations we examine the normalized density
of states, $\rho(\omega)$ and the onset of the pseudogap regime in
Fig.~\ref{fig:dos} for $\beta\varepsilon_{b}=0.47$ and temperature
$T/T_{{\rm F}}=0.2$, which corresponds to an interaction strength
$\eta=\ln[k_{{\rm F}}a_{{\rm 2D}}]\simeq1.5$. The density of states  is computed from
the spectral function $A(\mathbf{k},\omega)$,  which is found  by analytically continuing
the Green's function to real frequencies,
$A(\mathbf{k},\omega)=\text{Im}\, G(\mathbf{k},\omega+i0^{+})/\pi$.
This is achieved through Padè approximants \cite{Beach2000} and the
density of states then follows as the momentum average of the spectral
function $\rho(\omega)=\int d\mathbf{k}A(\mathbf{k},\omega)/(2\pi)^{2}$.
There are two methods used to calculate the density of states, analytically continuing 
 the self-energy or the Green's function directly. Using the self-energy produces 
 a smoother density of states as the numerical integration
of the spectral function is considerably simpler. This method is used for the calculation 
used in the $GG$ and $GG_0$ theory and the Green's function is directly continued
for the {\rm NSR} theory.

The interaction and temperature used in the calculation of $\rho(\omega)$
in Fig.~\ref{fig:dos} are experimentally attainable. At $T/T_{{\rm F}}=0.2$
and $\beta\varepsilon_{b}=0.47$ the converged chemical potential $\mu$ for each 
of the $T$-matrix theories is larger than zero, and for the $GG$ theory the 
compressibility is lowering. We see that for each of the $T$-matrix theories
the density of states at the chemical potential becomes suppressed,
and at either side we see an increase in the density of states, indicative 
of a pseudogap. There is no precise definition for the
onset of the pseudogap; it is, however, most likely too small an effect
in the $GG$ and $GG_{0}$ theories at this temperature and interaction
strength for us to confidently say that there is indeed a pseudogap,
however, at lower temperatures and larger interactions, the effect becomes
more pronounced \cite{Bauer2014}. Looking at the \textrm{NSR} theory 
there is a significant increase of the density of states and there is a pseudogap phase at
this temperature and interaction. Thus,  we would expect for 
an interaction strength of $\beta\varepsilon_{b}\simeq0.47$ and temperatures
lower than $T/T_{{\rm F}}\simeq0.2$ that the system would contain a pseudogap regime.

In conclusion, we have compared three $T$-matrix theories with experiment 
and found in the normal phase the fully-self consistent $GG$
$T$-matrix theory agrees well with a wide range of temperatures and
interactions. Comparatively, the $GG_{0}$ and \textrm{NSR} $T$-matrix
schemes underestimate the density and pressure equation of state in
the strongly interacting regime. Examining the density of states, we
have shown that each theory predicts a pseudogap
at a temperature and interaction strength accessible in current experiments.
Comparing the universal function $\tilde{\kappa}(\tilde{p})$ found from
experiment and the $T$-matrix theories, we see a difference at low temperature
close to the BKT transition. In order to understand the below $T_{c}$ thermodynamic properties,
a theory beyond the $T$-matrix approximations must be used where we can explicitly take into account
the superfluidity.

\textit{Note added:} Recently, we became aware of a related paper \cite{Boettcher2015} 
that examines the thermodynamics
of a 2D Fermi gas across the BCS-BEC crossover. This work found similar results 
but focused on the BEC side of the crossover.

\begin{acknowledgments}
We would like to thank Giacomo Bighin and Luca Salasnich us for giving their data,
 Meera Parish for useful discussions, and the
ARC Discovery Projects (FT130100815, DP140100637, DP140103231 and
FT140100003). 
\end{acknowledgments}

\bibliography{2DpaperV2}

\end{document}